\documentclass[11pt]{article}
\usepackage{a4wide}
\usepackage{xspace}
\usepackage{times}
\usepackage{epsfig}
\newcommand{\degree}{$^{\circ}$}

\newcommand{\Fref}[1]{Fig. {}\ref{#1}}
\newcommand{\fref}[1]{fig. {}\ref{#1}}
\newcommand{\etal}{{\it et al.}\@\xspace}
\newcommand{\A}{\@\xspace {\AA}\@\xspace}

\begin{document}
\title{Creation of multiple nanodots by single ions}
\maketitle
\begin{center}$^1$Ender Akc\"oltekin, $^1$Thorsten
Peters, $^1$Ralf Meyer, $^1$Andreas Duvenbeck, $^1$Miriam
Klusmann, $^2$Isabelle Monnet, $^2$Henning Lebius, $^1$Marika
Schleberger\footnote{Corresponding author:
marika.schleberger@uni-due.de}\\
 \vspace{.5cm} {\it $^1$ Universit\"at
Duisburg-Essen, FB Physik, 47048 Duisburg, Germany
\\
$^2$ CIRIL (CEA, CNRS, ENSICAEN), 14070 Caen Cedex 5, France}
\end{center}

{\bf  In the challenging search for tools that are able to modify
surfaces on the nanometer scale, heavy ions with energies of several
10 MeV are becoming more and more attractive. In contrast to slow
ions where nuclear stopping is important and the energy is
dissipated into a large volume in the crystal, in the high energy
regime the stopping is due to electronic excitations only
\cite{Bethe,Bloch,Lindhard}. Because of the extremely local
($\leq$ 1 nm) energy deposition with densities of up to
10$^{19}$W/cm$^2$, nanoscaled hillocks can be created under
normal incidence
\cite{Neumann,Bouffard,Khalfaoui}. Usually, each nanodot is due to
the impact of a single ion and the dots are randomly distributed. We
demonstrate that multiple periodically spaced dots separated by a
few 10 nanometers can be created by a single ion if the sample is
irradiated under grazing angles of incidence. By varying this angle
the number of dots can be controlled.}

As samples we used strontium titanate (SrTiO$_3$) single crystals, a
model substance for the class of perovskites (ABO$_3$). The
ferroelectric oxide SrTiO$_3$ is of use in a wide variety of
applications: SrTiO$_3$ is used in catalysis \cite{Catalysis} and is
discussed as a possible material for nuclear waste management
\cite{NuclearWaste}. SrTiO$_3$ is the insulator of choice for the
fabrication of future microelectronic devices. Because of its high
dielectric constant thin films of SrTiO$_3$ have only a very low
tunneling current compared to conventionally used SiO$_2$
\cite{Highk,Highk2}. In addition, it is widely used as substrate
material for GaAs solar cells and for the deposition of
high-temperature superconductors. For the latter, patterning of the
substrate is known to influence the critical current densities
\cite{FIB}. Therefore, the creation of regular nanometer sized
structures in SrTiO$_3$ might as well proof useful for technological
applications.

\begin{figure}[b]
\epsfig{figure=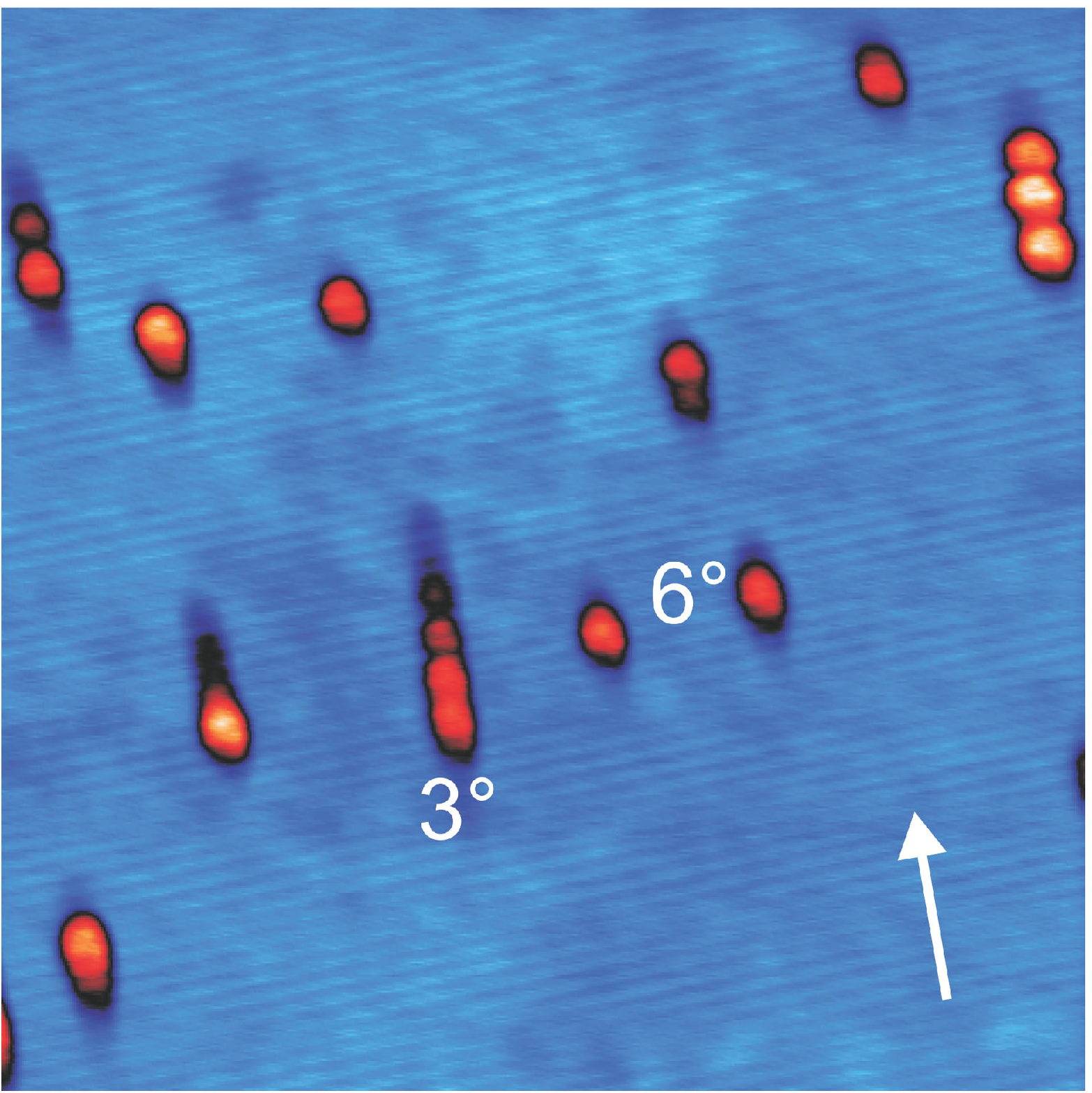,width=12cm} \caption{\label{Dots}
Typical nanodots on SrTiO$_3$ created by irradiation with 92 MeV
Xe ions. Frame size 500 x 500 nm$^2$, colour scale from 0 to 4.4
nm. The sample was irradiated two times. The angle of incidence
was $\Theta$~=~3\degree and $\Theta$~=~6\degree, respectively.
Each dot or chain, respectively, corresponds to a single ion hit.
To enhance the contrast false colouring was used. The arrow
indicates the direction of the incoming ions.}
\end{figure}

\begin{figure}
\epsfig{figure=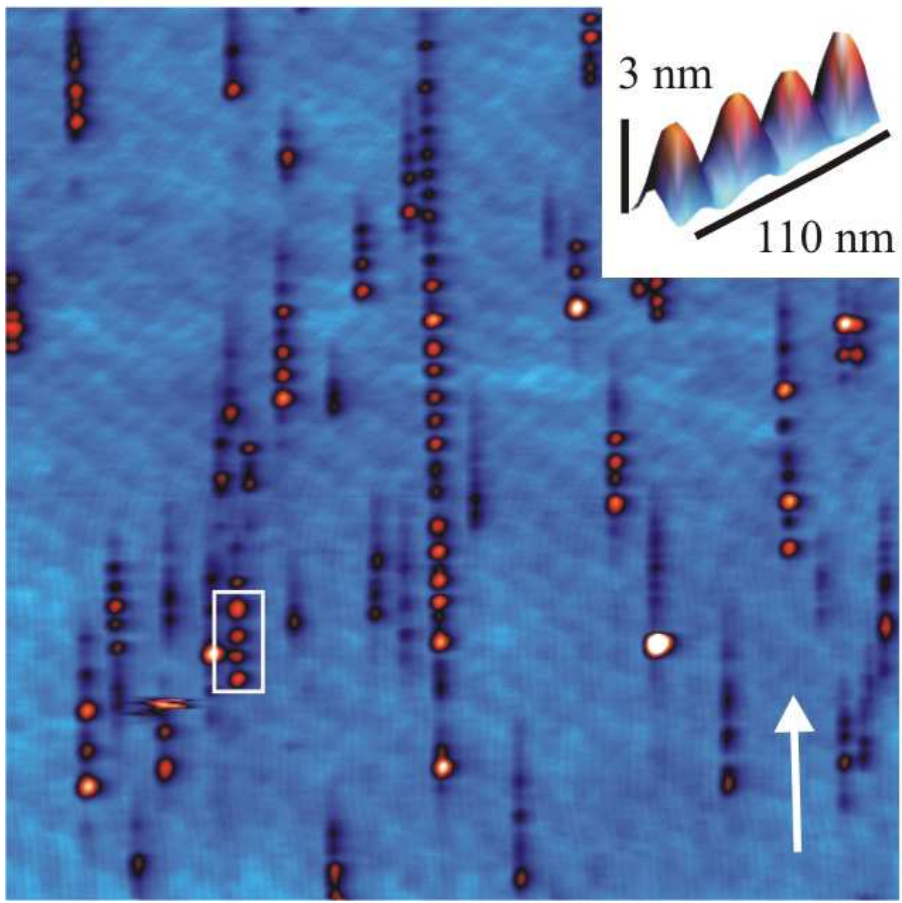,width=12cm} \caption{\label{Ketten}
Typical chains of nanodots on SrTiO$_3$ created by irradiation
with 92 MeV Xe ions. Frame size 1300 x 1300 nm$^2$, colour scale
from 0 to 3.7 nm. The sample was irradiated two times. The angle
of incidence was $\Theta$~=~1\degree and $\Theta$~=~2\degree,
respectively. The inset shows size and shape of the individual
dots within a chain.}
\end{figure}

\begin{figure}
\epsfig{figure=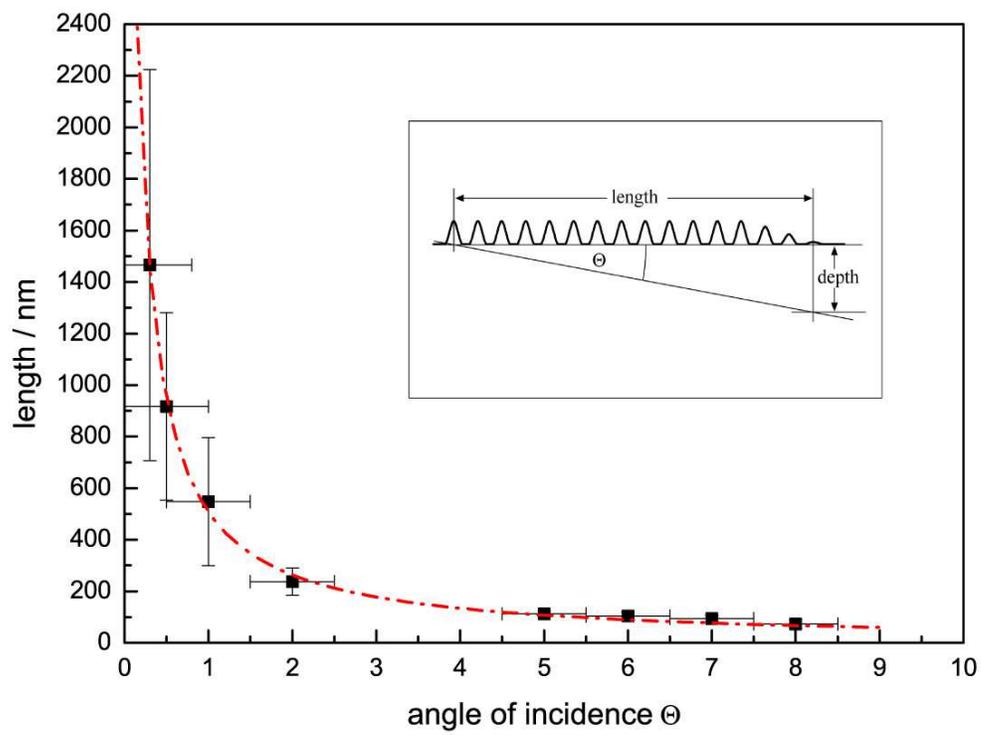,width=15cm} \caption{\label{Winkel}
Measured length $l$ of chains as a function of angle of
incidence. The dash-dotted line is a fit according to our model
with d~=~9.8 nm (see text).}
\end{figure}

\begin{figure}
\epsfig{figure=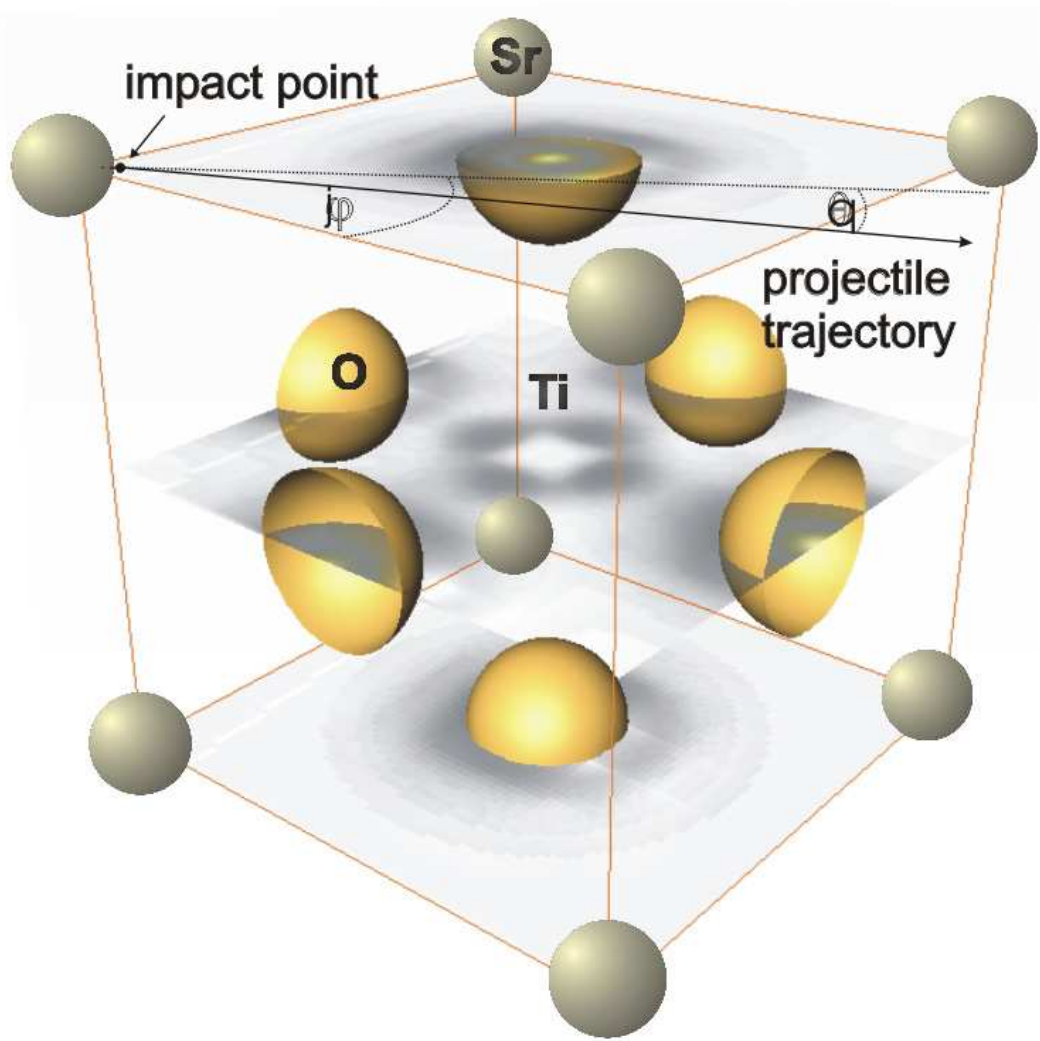,width=14cm}
\caption{\label{ElDensity} DFT calculation of the electron density
(gray shading) of SrTiO$_3$. Atom positions (besides the central
titanium atom) are also visualized. The arrow indicates a
possible projectile trajectory.}
\end{figure}

After irradiation under grazing angles of a few degrees with
respect to the surface plane (see \fref{ElDensity}) the samples
were imaged by atomic force microscopy (AFM) {\it in situ} under
ultra high vacuum (UHV) conditions. In \fref{Dots} we show an AFM
image of a sample that was irradiated under two different
incidence angles, resulting in two different types of defects. If
irradiated at $\Theta$~=~6\degree, slightly elongated hillocks
occur. If the angle is changed to $\Theta$~=~3\degree, round
almost evenly spaced hillocks appear which are still overlapping.
At angles smaller than $\Theta$~=~3\degree~ the dots are clearly
separated and the tracks look like pearls on a chain as can be
seen in \fref{Ketten} (this sample was irradiated at
$\Theta$~=~1\degree and at $\Theta$~=~2\degree). The number of
chains corresponds to the number of ions. This means each chain
containing dozens of nanodots is created by a single ion
traveling through the solid. From analyzing more than 600
individual chains irradiated under different angles $\Theta$ we
find that the length $l$ of the chains can be controlled by
varying the angle of incidence as can be seen in \fref{Winkel}.
At $\Theta$~=~1\degree most of the chains are already about half
a micron in length. The average height value we find for all
chains stemming from irradiations at incidence angles of less
than 3\degree is (2.5 $\pm$ 1.0) nm. They are of circular shape
with a diameter (FWHM) of (25 $\pm$ 5) nm. The distance between
individual hillocks depends on the angle of incidence and is e.g.
(35 $\pm$ 3) nm for the marked chain in \fref{Ketten}. Similar
measurements (not shown here) on irradiated TiO$_2$ and
Al$_2$O$_3$ show the same chains of dots as on SrTiO$_3$. We take
this as an indication that this phenomenon may be typical not
only for perovskites but also for other oxidic materials.

To date, the patterning of ferroelectric oxides has been achieved
by self-organizing (\lq bottom-up\rq) approaches
\cite{Zhang,Szafraniak} as well as by \lq top-down\rq~ methods
such as lithography using low energy ions \cite{Habermeier},
evaporation through a shadow mask \cite{Shin} or a combination of
both \cite{Ruzmetov}. With these techniques the created
structures are either randomly distributed or shaped, or the
periodicity is limited to 100 nm at best. To our knowledge, no
other technique that is able to create regular shaped nanodots on
a perovskite surface with a spacing of 35 nm has so far been
established.

In order to describe the ion-solid interaction the so-called
thermal spike (TS)-model has been developed \cite{Seitz}.
Toulemonde \etal applied it to the special case of swift heavy
ions where electronic stopping dominates over nuclear stopping by
orders of magnitude \cite{Toulemonde}. It describes the formation
of cylindrical tracks of amorphous or recrystallized material
inside the solid. It does neither explain the formation mechanism
of hillocks on the surface nor the formation of hillock chains.
The starting point to gain an understanding of the formation of
nanodot chains on SrTiO$_3$ are early experiments with van der
Waals materials such as MoS$_2$ or WSe$_2$. In these materials
intermittent fission tracks were observed by transmission
electron microscopy \cite{Chadderton}. These discontinuous tracks
were explained by the strong spatial anisotropy of a van der
Waals crystal lattice. The $\pi$- electrons are fully contained
within the crystal planes and there is only a weak hydrogen bond
between the planes. Each time the projectile hits a lattice plane
with a high density of electrons, enhanced electronic stopping
occurs. This significant anisotropy leads to periodic spikes of
radiation damage in this type of materials as observed by Morgan
and Chadderton \cite{Chadderton}. Such a triggered loss process
could also explain the distinct features we observe.

If the ion travels through SrTiO$_3$ it will continuously
interact with the electronic system of the material but energy
loss is more probable where the electron density is high. There,
the sudden interaction of the projectile with the electronic
system gives rise to a sharply peaked energy distribution. How
this electronic excitation finally leads to the modification of
the material (craters, hillocks, tracks, etc.) is still not
exactly known. The thermal spike model describes the energy
transport out of the electronically heated region, whereas the
Coulomb explosion model \cite{Fleischer} couples the electronic
excitation to atomic motion caused by the repulsive forces acting
in the transiently ionized region. Both models cover important
aspects of the creation processes for material modifications
\cite{Klaumuenzer}. In our experiment the strong ionization rate
of the swift heavy ion leads to a significant charge imbalance.
From Shimas empirical formula \cite{Shima} and extrapolating
experimental data \cite{Shima2}, respectively, we find that Xe
ions moving through SrTiO$_3$ with 92 MeV have an effective
equilibrium charge close to the original charge state of
$q_{\mathrm{eff}}$=23. However, our data does not allow us to
make a clear correlation with neither of the two scenarios. In
any case, for both models to be predictive the initial conditions
need to be known. A strong anisotropy of the electronic structure
as e.g. in a van der Waals material will influence the loss
process significantly, be it connected either to a Coulomb
explosion or to a thermal spike.

\begin{figure}
\epsfig{figure=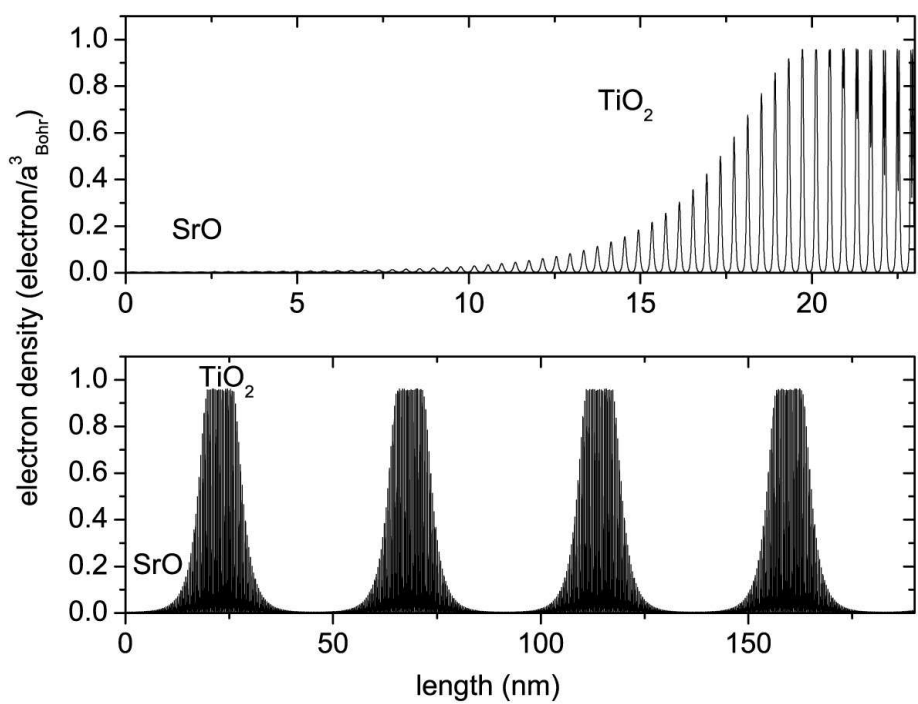,width=14cm} \caption{\label{GeoModell}
Simulation of the electron density which the ion encounters when
traveling through the crystal lattice at a grazing angle of
$\Theta$~=~0.5\degree with respect to the (100)-surface plane of
a SrTiO$_3$ single crystal (see \fref{ElDensity}).}
\end{figure}

Note however, that SrTiO$_3$ is not a van der Waals material and the
observed features cannot be simply attributed to the ion crossing
crystal layers of a homogeneous electron density. If the anisotropy
was exclusively parallel to the surface (as is the case with
MoS$_2$), we would observe that the number of nanodots was constant
and the distance between dots would vary as a function of the angle
of incidence. If the anisotropy was exclusively normal to the
surface, we would observe a varying number of dots, but the distance
between them would be constant. In our case the anisotropy is given
parallel as well as normal to the surface. Therefore, the observed
distance between dots and the number of the dots as a result of a
triggered loss process needs to be discussed taking the full three
dimensional electronic structure into account. To this end, we
performed {\it ab-initio} density functional theory (DFT)
calculations to determine the electronic density of SrTiO$_3$ and
projected the density onto the plane of the traveling ion (see
\fref{GeoModell}).

Most of the electrons are located around the oxygen atoms and the
density is higher in the TiO$_2$ planes than in the SrO planes
(see \fref{ElDensity}). Simple geometrical considerations taking
the electron density into account yield a very good qualitative
agreement with the periodic defects occuring in SrTiO$_3$ as can
be seen from \fref{GeoModell}.
In this approach we have used the DFT data to calculate the
electron density encountered by an ion following a typical
trajectory (azimuthal angle $\varphi=0$\degree, see
\fref{ElDensity}) through a perfect SrTiO$_3$ crystal under a
grazing angle of $\Theta=0.5$\degree. On its way through a single
lattice plane the ion interacts with the electrons of several
dozens of oxygen atoms. Each peak in the fine structure (see
upper panel in \fref{GeoModell}) is due to the interaction of the
ion with the electrons surrounding one oxygen atom. The sum of
these interactions within one crystal plane (large peaks in the
lower panel of \fref{GeoModell}) creates a strong local
excitation leading finally to a nanodot on the surface. The ion
then typically travels several hundred \AA ngstr\"om through the
crystal without energy loss before it again encounters an area
with a high enough electron density.

Even though the exact mechanisms of the observed dot formation
(\fref{Dots} and \fref{Ketten}) are not known, we may assume the
space- and time dependent \textit{generation} of excitation energy
$E_{s}(\vec{r},t)$ along the trajectory of the projectile
$\vec{r}_{p}(t)$ as well as the \textit{transport} of excitation
energy to play a key role in that process. In the energy regime
considered here, the generation of excitation energy can be
approximately treated within the frame of the Lindhard model
\cite{Lindhard2} of electronic stopping. Employing that framework,
the excitation energy $dE_s(\vec{r}_{p}(t))$ that is transferred
from the kinetic energy of the projectile into the electronic
system within the time interval $dt$ is proportional to the
momentary kinetic energy of the projectile and - even more
important in our context - to the local electron density
$n_{el}(\vec{r}_{p}(t))$ provided by the ab-initio DFT
calculations as explained above. We assume the time evolution of
the four-dimensional profile $E(\vec{r},t)$ of excitation energy
within the solid to be described by the diffusion equation
\begin{equation}
\frac{\partial E}{\partial t}(\vec{r},t) - D \nabla^2 E(\vec{r},t)
=
\frac{dE_{s}}{dt}(\vec{r}_{p}(t),E_{kin}(\vec{r}_{p}(t)),n_{el}(\vec{r}_{p}(t)))
\label{eq:1}
\end{equation}
with $D$ denoting the diffusion coefficient. The projectile
trajectory $\vec{r}_{p}(t)$ entering Eq.~(\ref{eq:1}) is obtained by
numerically integrating the Newtonian equations of motion including
an effective friction term originating from the Lindhard treatment.
In view of the fact that on the one hand the lateral dimensions of
the observed nanodots are of the order of several tens of
nanometers, but, on the other hand, the electron density within one
unit cell varies on the sub-{\A}ngstr\"om length-scale, a
straight-forward numerical treatment of Eq.~(\ref{eq:1}) is hampered
by the complexity of the problem.

It should be emphasized here that in our case (i) the consideration
of the local electron density on a sub-{\A}ngstr\"om length scale
and (ii) the necessity for the correct incorporation of the surface
plane into the excitation energy transport process principally
disallows the reduction of the problem to radial symmetry with
respect to the swift heavy ion track as usually done in (TS)-model
calculations \cite{Toulemonde2}. In contrast to the (TS)-model
calculations, we propose to perform a nested two-step calculation
approach as follows: In the first step, we take advantage of the
straightness of the ion track and - for given impact parameters
$(\Theta,\phi)$ - geometrically determine the set of intersection
points through the unit cells along the trajectory of the
projectile. This procedure is performed for a total trajectory
length $L$ which corresponds to a laterally projected range larger
than the observed period length of the nanodots. Starting from the
unit cell at the impact point, each traversed cell is discretized
into more than 10$^{5}$ voxels to match the sub-{\A}ngstr\"om
resolution of $n_{el}(\vec{r},t)$. A numerical integration of the
equations of motion of the projectile employing the highly-resolved
electron density yields the total generated excitation energy
$dE_s(\vec{r}_{p}(t))$ as well as the traveling time $dt$ for that
cell. Naturally, the kinetic energy of the heavy ion is reduced
after traversing one cell due to electronic stopping.

This reduced energy as well as the exit point of one cell are
taken as the new initial parameters for the analogous calculation
carried out for the subsequent cell. Iteration of this algorithm
yields a set of $(dE/dt)_{i}$ which in the second step of our
model, are treated as distinct point sources of excitation energy
in a discrete representation of Eq.~(\ref{eq:1}) with a
grid-spacing in units of one elementary cell. Using that coarse
discretization the numerical solution of Eq.~(\ref{eq:1}) can be
obtained via a Green\lq s functions approach for the half space
diffusion problem.

\begin{figure}
\epsfig{figure=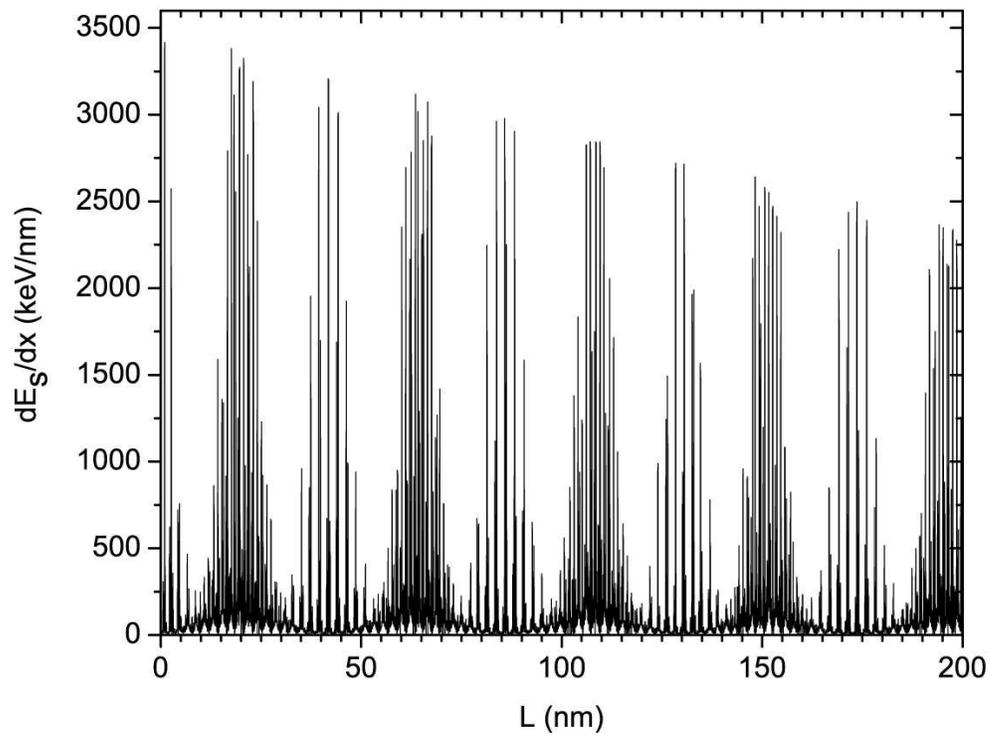,width=15cm}
\caption{\label{Energyspike} Electronic stopping $dE_{s}/dx$
along the track of a Xe projectile hitting a SrTiO$_3$ single
crystal with $\Theta$~=~0.5\degree and $\phi$~=~10\degree.}
\end{figure}

In the present study we focus our interest on the striking
periodicity of the nanodots which should already be present in the
space dependence of the electronic stopping (ES) $dE_{s}/dx$ along
the trajectory derived from $\frac{dE_{s}}{dt}(\vec{r}_{p}(t))$ of
Eq.~(\ref{eq:1}). \Fref{Energyspike} shows the result of an
exemplary calculation of this ES for a 0.711 MeV/u Xe projectile
hitting a SrTiO$_3$ crystal under $\Theta=0.5$\degree and
$\phi=10$\degree. Two kinds of peaks can be seen, one originating
from the TiO$_2$ planes, the second one from the SrO planes. The
average ES in the TiO$_2$ planes
is higher by a factor of 2 (integrating over the peak area).
Thus, for this particular choice of impact parameters, the
periodicity of the dot formation may be governed by the
contribution of the TiO$_2$ planes. The decrease of the maximum
peak heights with increasing track length $L$ due to electronic
friction will be one factor limiting the total chain length $l$.
A more quantitative discussion of features such as e.g. chain
length, track radius and temporal dynamics goes beyond the scope
of this paper because it would require a detailed analysis of the
full four-dimensional excitation energy profile $E(\vec{r},t)$
obtained from Eq.~(\ref{eq:1}) using the ES discussed above.

Our model is further corroborated by plotting the measured length
of the chains as a function of the angle of incidence as shown in
\fref{Winkel}. The data can be fitted nicely by using
$l(\Theta)=d/(tan \Theta)$. Here, $l$ is the length of the chain,
and $d$ is the maximum depth from where the excitation starts (see
inset in \fref{Winkel}). In addition, we frequently observe, that
at the very end of a chain the height of the dots decreases
monotonic. These findings could be explained if we assume that
the ion is traveling already too deep below the surface and the
material modifications do not reach the surface any longer. In
this way, we can determine the radius of the modified volume from
our data. Our value of ca. 10 nm is larger by a factor of 2 than
the effective latent track radii derived from irradiation
experiments on different insulators under 90\degree
\cite{Meftah,Szenes}. Repeating our experiment with an amorphous
layer of SiO$_2$, we find periodic dots as well \cite{Carvalho}.
It is not clear yet, what the origin of this periodicity is, but
it could be the electron density, maybe even of the underlying
crystalline Si substrate.

In conclusion, we have demonstrated how to produce periodic
nanodots on oxidic surfaces by a single ion hit. We propose that
the anisotropic electron density of the material gives rise to a
triggered energy loss process. The resulting nanodots imaged by
AFM thus represent a direct view of the projection of the
three-dimensional electronic density onto the surface.
\\
\\
\clearpage
 {\small
\section*{Methods}
\subsection*{Experimental setup, sample preparation and image processing}
The experiments were performed at the beamline IRRSUD of the Grand
Accelerateur National d'Ions Lourds (GANIL) in Caen, France. An
UHV-AFM/STM (Omicron Nanotechnology, Taunusstein) was mounted
directly to the chamber where the irradiation with swift heavy ions
took place. The base pressure was 2 $\times 10^{-8}$ mbar in the
differentially pumped irradiation chamber and 1 $\times 10^{-10}$
mbar in the AFM. As samples we used commercially available
SrTiO$_3$(100), TiO$_2$(100) and Al$_2$O$_3$(1102) crystals
(CrysTec, Berlin). The samples were cleaned with acetone. We did not
use any etching to get rid of excess SrO terminations since this
would not play a role in the hillock formation. AFM images taken
before irradiation to check the cleanliness of the samples
frequently showed atomically flat terraces separated by atomic
steps.

The samples were irradiated with a beam of Xe$^{23+}$ ions at
0.711 MeV/u scanned over the whole surface area of the target.
Typically, a fluence of 1$\times 10^{9} ~\rm ions/cm^2$ was chosen
resulting in 10 impacts per $\mu$m$^2$ on average. At this fluence
enough events are produced to obtain a good statistic but the
probability that two ions hit the same spot resulting in
non-linear phenomena is still sufficiently low. For SrTiO$_3$ the
chosen projectile/energy combination results in an energy loss of
19.1 keV/nm as calculated with SRIM \cite{SRIM}. The angle of
incidence with respect to the surface was varied between
$\Theta$=1\degree and $\Theta$=6\degree. Due to the experimental
setup the uncertainty of the absolute angle is not better than
$\mathrm{\pm 1}$\degree. Due to technical limitations the
azimuthal angle $\varphi$ was not controlled in our experiment.
Immediately after irradiation all samples were measured with the
AFM using the contact ({\it F} = 0.2 nN) as well as the dynamic
(non-contact) mode ({\it df} = -10 Hz) {\it in situ}. Additional
data from {\it ex-situ} measurements was used for \fref{Winkel}.
No chemical etching or post-irradiation treatment was applied.
This ensures that all observed topographical features can be
attributed unambiguously to the irradiation.

All AFM images were recorded with the Omicron SCALA software
version 4.1 and processed with the Nanotec Electronica SL WSxM
software, version 4.0 Develop 8.3. From the raw data (400
$\times$ 400 data points) only a plane was subtracted. The colour
code was changed using the palette {\it flow.lut}. No change of
contrast (1) or brightness (0) was used.

\subsection*{DFT calculations} The density functional theory
(DFT) calculations were performed using the ABINIT package
\cite{abinit} together with pseudopotentials generated by the
fhi98pp code \cite{fhi98pp} in order to determine the distribution
of the electrons in $\mathrm{SrTiO}_3$. For the exchange-correlation
energy the Perdew-Burke-Ernzerhof generalized-gradient approximation
functional \cite{PBE} was used. In a first step, the equilibrium
lattice constant of the investigated system was obtained. The result
is 7.53\,$a_\mathrm{B}$ for $\mathrm{SrTiO}_3$. Subsequently the
electron density at the equilibrium lattice constant was derived. A
common kinetic-energy cutoff-energy of 96\,Hartree for the expansion
of the wave functions and a $8\times 8\times 8$ k-point mesh were
used in all calculations.
\\
\\
\section*{Acknowledgement}
Financial support by the DFG - SFB616: {\it Energy dissipation at
surfaces} and SFB 445: {\it Nanoparticles from the Gas Phase}, by
the GANIL (Project S18) and by the European Community FP6 -
Structuring the ERA - Integrated Infrastructure
Initiative-contract EURONS n$\mathrm{^o}$ RII-CT-2004-506065 is
gratefully acknowledged. We thank Philippe and Frederic JeanJean
for their help with the experiment and Andreas Reichert for
valuable discussions. }

\newpage

\bibliographystyle{unsrt}
\bibliography{Iondots}

\newpage

\end{document}